\documentclass[12pt]{article} 
\usepackage{graphicx}
\usepackage{amsfonts}
\usepackage{epstopdf}
\date{}

\newtheorem{prop}{Proposition}\newtheorem{defin}{Definition}

\vfill\title{Chain Rotations: a New Look at Tree Distance
}
\author {Fabrizio Luccio\thanks{Dipartimento di Informatica, Universit\`{a} diPisa, luccio@di.unipi.it}\and Linda Pagli\thanks{Dipartimento di Informatica, Universit\`{a} di Pisa, pagli@di.unipi.it}} 

\begin{document}
\maketitle
\begin{abstract}As well known the rotation distance $D(S,T)$ between two binary trees $S$, $T$ of $n$ vertices is the minimum number of rotations of pairs of vertices to transform $S$ into $T$. We introduce the new operation of {\em chain rotation} on a tree, involving two chains of vertices, that requires changing exactly three pointers in the data structure as for a standard rotation, and define the corresponding {\em chain distance} $C(S,T)$. As for $D(S,T)$, no polynomial time algorithm to compute $C(S,T)$ is known. We prove a constructive upper bound and an analytical lower bound on $C(S,T)$ based on the number of maximal chains in the two trees. 
In terms of $n$ we prove the general upper bound $C(S,T)\leq n-1$ and we show that there are pairs of trees for which this bound is tight. 
No similar result is known for $D(S,T)$ where the best upper and lower bounds are $2n-6$ and $\frac{5}{3}n-4$ respectively.
\end{abstract}


\section{A new definition of tree distance}
Consider a rooted binary tree $T$ of $n$ vertices identified with the integers from 1 to $n$ in infix order as for a binary search tree (in the following the term {\em tree}  always refers to trees of this form). 
A {\em subtree} $W$ of $T$ is a tree rooted in a vertex $v$ of $T$ and containing all the descendants of $v$ in $T$. The vertices of $W$ correspond to an integer interval, e.g., in the tree $T$ of Figure 1 the subtree rooted at 7 corresponds to the interval [4,8].
A {\em rotation} of two adjacent vertices is an operation preserving the infix order through the change of three pointers. In Figure 1 the rotation between vertices 5 and 7 produces a tree $T'$ where the right pointers of 3 and 5, and the left pointer of 7, have been changed. The inverse rotation between 7 and 5 transforms $T'$ into $T$.

Rotations were originally defined to keep binary search trees balanced. In~\cite{CW82} Culik and Wood have defined the {\em rotation distance} $D(S,T)$ between two trees $S$ and $T$ as the minimum number of rotations needed to transform $S$  into $T$.
$D(S,T)$ has then been adopted in combinatorics as a standard measure of distance between trees, and has a role in computational biology where a comparison between evolutionary trees is done on the basis of subtree transfer~\cite{D+98}.
A transformation requiring $D(S,T)$ rotations is called {\em optimal}. 

A constructive upper bound $D(S,T)\leq 2n-2$ was given in~\cite{CW82} and was improved to $2n-6$ in the seminal work of Sleator, Tarjan, and Thurston~\cite{STT86} where the authors transformed the problem into polygon transformation via diagonal flips. Since then a rich literature has appeared on the subject, nevertheless no efficient algorithm has been proposed to determine $D(S,T)$, nor it is known whether the problem is NP-hard. In particular interesting estimates for $D(S,T)$ have been given in~\cite{P00,R99}, a significant approximation algorithm has been proposed in~\cite{D+98}, and other works have been directed to establish significant lower bounds~\cite{D10,LMP10}. All in all rotation distance is a classical topic and has led to many elegant results.

\begin{figure}\label{rot-def}
\begin{center}
\includegraphics[scale=0.8]{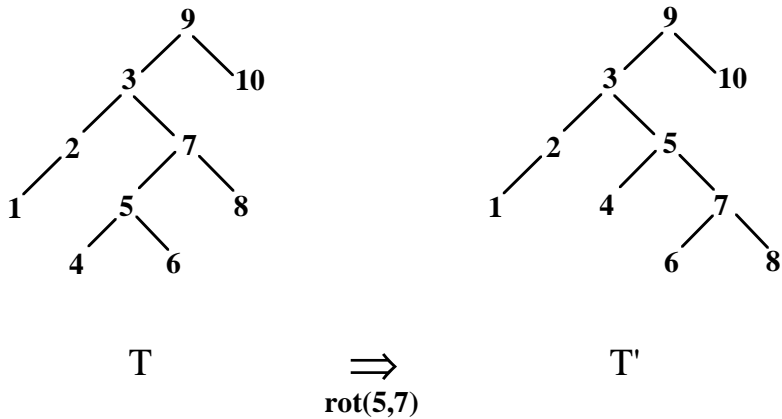}
\caption{A rooted binary tree $T$ and the effect of a vertex rotation.}\end{center}
\end{figure}


An important concept is the one of {\em equivalent edges}, that is pairs of edges, one in $S$ and one in $T$, whose deletion splits both $S$ and $T$ in two parts $S_1,S_2$ and $T_1,T_2$ where $S_1,T_1$ are subtrees of $S,T$ containing the same subset of vertices (hence correspond to the same integer interval), and $S_2,T_2$ containing the remaining vertices. 
In Figure 1 the edges (3,7) of $T$ and (3,5) of $T'$ are equivalent, with the resulting subtrees of $T$ and $T'$ respectively rooted in 7 and 5 and corresponding to the interval [4,8]. The remaining portions of $T,T'$ contain vertices 1, 2, 3, 9, 10.
If a pair of equivalent edges exists, any optimal
transformation of $S$ into $T$ can be done independently on $S_1,T_1$ and $S_2,T_2$. Note that the equivalent edges can be determined in linear time.
Letting $e$ denote the number of pairs of equivalent edges,
the two trees are split accordingly into $e+1$ pairs of trees to be processed independently. It has been proved that, for $e=0$, at least $n-1$ rotations are needed to transform $S$ into $T$~\cite{CSJ10,LMP10}, then the lower bound $D(S,T)\geq n-e-1$ follows. 
We assume that the splitting of $S,T$ has been done beforehand, then we shall work on trees without equivalent edges. 

Rotations were defined in the 
field of data structures with the unquestionable merit of being local operations that require exactly three pointer changes. This implies the replacement of two vertices (5 and 7 in Figure 1) and one subtree transfer (subtree [6], here composed of one vertex, migrates from right subtree of 5 to left subtree of 7).
We now propose a more general operation called {\em c-rotation}, where $c$ stands for {\em chain}, that is done on chains instead of single vertices and also requires three pointer changes and one subtree transfer. A standard rotation is a special case of c-rotation if the chain contains only one vertex. We let:

\vspace{2mm}
\noindent {\bf Terminology and notation}. A {\em left chain} [$u$-$v$], $u>v$, in a tree $T$ is a sequence of vertices connected to one another with left pointers, from $u$ (the highest) to $v$ (the lowest). 
A {\em maximal left chain} is such that no other left chain contains it. 
The {\em complete left chain} [$n$-1] contains all the vertices linked with left pointers in the order $n$, $n$-1, ..., 1.
A {\em right chain}, a {\em maximal right chain}, and the {\em complete right chain} [1-$n$] are similarly defined.
A left or right chain containing only one vertex $u$ is denoted by [$u$].
$L_T$ and $R_T$ respectively denote the number of maximal left chains and of maximal right chains in $T$.
\vspace{2mm}

In the tree $T$ of Figure 1: [7-4] is a maximal left chain; [5-4] is a not maximal left chain, being contained in [7-4]; [5-6] and [4] are maximal right chains. We have $L_T=5$ and $R_T=6$. We have:

\begin{prop}\label{prop-LR}In a tree $T$ of $n$ vertices we have $L_T+R_T=n+1$.
\end{prop}

Proposition~\ref{prop-LR} is proved by simple induction on $n$. The basis is $n=1$, for which we have $L_T=1$ and $R_T=1$. Letting the proposition to be true for $n-1$, insert a new leaf $v$ in $T$ as a child of an existing vertex $u$. If $v$ is a left child of $u$, $L_T$ is unchanged and $R_T$ is increased by 1. If $v$ is a right child of $u$, $R_T$ is unchanged and $L_T$ is increased by 1. Then the proposition is true for $n$. Similarly note that $L_T$ and $R_T$ are respectively equal to the numbers of non-null right and left pointers in $T$ plus 1. As a consequence the values of $L_T,R_T$ can be computed in $O(n)$ time in a tree traversal.  

As for standard rotations, a c-rotation can be inverted. If needed we shall distinguish between direct and inverse c-rotations, defined as follows:

\begin{defin}\label{def-rot}A {\em (direct) c-rotation rot}$([u$-$v], w)$ in a tree $T$, where $[u$-$v]$ is a left chain and $u$ is the right child of $w$, is a local operation where: (i) $u$ takes the place of 
$w$ (i.e. $u$ becomes a child of the parent $x$ of $w$, if any); (ii) $w$ becomes the left child of $v$; and (iii) the left subtree of $v$, if any (i.e., if  $[u$-$v]$ is not maximal), becomes the right subtree of $w$. The definition also holds for a right chain $[u$-$v]$ exchanging the terms ``left'' and ``right'' whenever they occur.
\end{defin}

In the tree $T$ now repeated in Figure 2, the c-rotation rot([7-5],3) produces the tree $T''$. Note that a direct c-rotation merges two chains into one, and [$u$-$v$] may be a maximal or a non maximal chain. 

\begin{defin}\label{def-inv rot}
 An {\em (inverse) c-rotation rot}$(w$,$[u$-$v])$ in a tree $T$, where $[u$-$v]$ is a left chain and $w$ is left child of $v$ (then $w$ is in the same left chain of $[u$-$v]$), 
is a local operation where: (i) $w$ takes the place of $u$ 
(i.e. $w$ becomes a child of the parent $x$ of $u$, if any) and $u$ becomes the right child of $w$; and (ii) the right subtree of $w$, if any, becomes the left subtree of $v$.
Again the definition also holds for a right chain $[u$-$v]$ exchanging the terms ``left'' and ``right'' whenever they occur.
\end{defin}

In Figure 2 the inverse c-rotation rot(3,[7-5]) in $T''$ produces the tree $T$ again. Note that an inverse rotation splits a chain in two, and the chain [$u$-$v$] cannot be maximal. We immediately have:

\begin{figure}
\begin{center}
\includegraphics[scale=0.8]{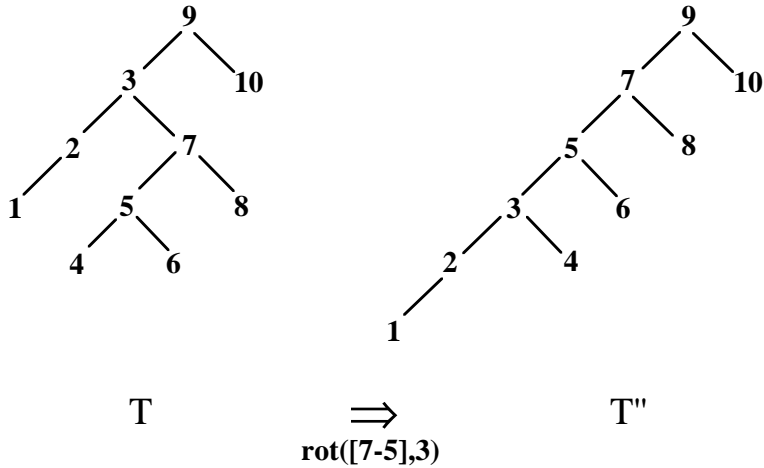}
\caption{A direct c-rotation on the tree $T$ of Figure 1 merging the chains [7-5] with [9-1].}
\end{center}
\end{figure}

\begin{prop}\label{prop-pointers}
In a (direct or inverse) c-rotation three pointers change. Namely, for {\em rot}$([u$-$v], w)$, where $x$ is the parent of $w$ if any,  we change: (i) the pointer from $x$  to $w$ (or the outside pointer to the tree if $w$ is the tree root); (ii) the left (respectively right) pointer of $v$; (iii) the right (respectively left) pointer of $w$. For {\em rot}$(w$,$[u$-$v])$ we change: (i) the pointer from $x$ to $u$ (or the outside pointer to the tree if $u$ is the tree root); (ii) the right (respectively left) pointer of $w$; (iii) the left (respectively right) pointer of $v$.  
\end{prop}

In rot([7-5],3) of Figure 2 the left pointer of the parent $x=9$ of $w=3$ now points to $u=7$; the left pointer of $v=5$ now points to $w=3$; the right pointer of $w=3$ now points to the left child 4 of $v=5$. The effect of the inverse rotation rot(3,[7-5]) is specular. Note that if [$u$-$v$] is a maximal chain $v$ has no left (respectively right) child, then in a direct rotation rot$([u$-$v], w)$ the right (respectively left) pointer of $w$ becomes null.

\begin{defin}\label{def-C}Given two trees $S,T$ of $n$ vertices, the {\em chain distance} 
$C(S,T)$ is the minimum number of c-rotations needed to transform $S$ into $T$.
\end{defin}

As a trivial example we have $C(T,T'')=1$ in Figure 2, however, determining the chain distance in the general case is a hard problem. 


\section{Upper and lower bounds on $C(S,T)$}\label{bounds}

We start giving a transformation algorithm between two trees $S,T$ based on c-rotations. The possibility of inverting a c-rotation suggests a strategy often used for regular rotations, e.g. see~\cite{LP89}. 
First it is decided how to transform both $S$ and $T$ into a proper {\em target tree} $Z$. Then the overall procedure will be $S\rightarrow Z\rightarrow T$, where $Z\rightarrow T$ is done by inverting the c-rotations of $T\rightarrow Z$ and applying them in opposite order. $C(S,T)$ is upper bounded by the sum of c-rotations in $S\rightarrow Z$ and $T\rightarrow Z$.  
The target tree chosen here is either the complete left chain [$n$-1] or the complete right chain [1-$n$]. 

In Figure 3 we show the structure of two algorithms for transforming a tree $Y$ of $n$ vertices into the chain [$n$-1] (ROTLEFT), or into the chain [1-$n$] (ROTRIGHT). Clearly both algorithms can be implemented to run in linear time. We have:

\begin{prop}\label{prop-UB}
$C(S,T)\leq$ {\em min(}$L_S+L_T-2, R_S+R_T-2\,${\em)}.
\end{prop}

Proposition~\ref{prop-UB} has a simple constructive proof. The transformation $S\rightarrow T$ can be performed as a combination of $S\rightarrow [n$-1] and $T\rightarrow [n$-1], or as a combination of $S\rightarrow$ [1-$n$] and $T\rightarrow$ [1-$n$], using ROTLEFT or ROTRIGHT. 
Since the number of c-rotations executed by ROTLEFT($Y$) and by ROTRIGHT($Y$) are $L_Y-1$ and $R_Y-1$ respectively, the proposition follows. Since ROTLEFT and ROTRIGHT use only direct c-rotations, 
the overall transformation  $S\rightarrow Z\rightarrow T$ will consists of direct c-rotations in $S\rightarrow Z$ and inverse c-rotations in  $Z\rightarrow T$. We can now derive an upper bound on $C(S,T)$ as a function of $n$, namely:

\begin{figure}
\small{
{\bf algorithm} ROTLEFT($Y$):
\vspace{1mm}

\hspace{5mm} {\bf while} ($\,\exists$ more than one left chain in $Y$) 
\vspace{1mm}

\hspace{12mm} merge a maximal left chain [$u$-$v$] with the chain containing the parent $w$ of $u$ 
\vspace{1mm}

\hspace{19mm}by applying rot([$u$-$v$],$w$). 

\vspace{3mm}
{\bf algorithm} ROTRIGHT($Y$):
\vspace{1mm}

\hspace{5mm} {\bf while} ($\,\exists$ more than one right chain in $Y$) 
\vspace{1mm}

\hspace{12mm} merge a maximal right chain [$u$-$v$] with the chain containing the parent $w$ of $u$ 
\vspace{1mm}

\hspace{19mm}by applying rot([$u$-$v$],$w$). 
}
\begin{center}
\caption{The algorithms ROTLEFT and ROTRIGHT for transforming a tree $Y$ into a complete chain. }
\end{center}
\end{figure}


\begin{prop}\label{prop-UBn}
$C(S,T)\leq n-1$.

\noindent{\bf Proof.} {\em For proving an upper bound valid for all trees we must compute the maximum of the function {\em f = min}($L_S+L_T-2, R_S+R_T-2$) given in Proposition~\ref{prop-UB} under the variation of the parameters involved. Recalling from Proposition~\ref{prop-LR} that $L_T+R_T=n+1$ for any tree $T$ and letting $L_S+L_T=\alpha$, we can reformulate the function as {\em f = min}($\alpha-2, 2n-\alpha$) with $\alpha$ growing linearly in the range $2\leq \alpha \leq 2n$. It is now easy to prove that $f$ is maximized for $\alpha = n+1$ and the given bound follows.
}  \hfill Q.E.D.
\end{prop}

For proving a lower bound on $C(S,T)$ we must rely on the properties of c-rotations. Working on the maximal chains of $S,T$ we have:

\begin{prop}\label{prop-LB1}
$C(S,T)\geq |L_S-L_T|$.

\noindent{\bf Proof.} {\em In a tree $Y$ a direct c-rotation rot([$u$-$v$],$w$), where [$u$-$v$] is a maximal left (respectively right) chain, induces the changes $L_Y=L_Y-1$ and $R_Y=R_Y+1$ (respectively $L_Y=L_Y+1$ and $R_Y=R_Y-1$). If instead [$u$-$v$] is not maximal the values of $L_Y$ and $R_Y$ remain unchanged (for example see the c-rotation in Figure 2). An inverse rotation rot($w$,[$u$-$v$]), where [$u$-$v$] is a left (respectively right) chain and $w$ has no right (respectively left) child, induces the changes $L_Y=L_Y+1$ and $R_Y=R_Y-1$ (respectively $L_Y=L_Y-1$ and $R_Y=R_Y+1$). If instead $w$ has a right (respectively left) child the values of $L_Y$ and $R_Y$ remain unchanged (for example invert the rotation in Figure 2). Then a c-rotation may change the value of $L_Y$ by at most one. The given bound immediately follows because, after the transformation of $S$ into $T$, the two trees must have the same number of maximal left chains. 
}  \hfill Q.E.D.
\end{prop}

Note that from Proposition~\ref{prop-LR} we have $L_S=n+1-R_S$ and $L_T=n+1-R_T$, hence $L_S - L_T=R_T-R_S$. This implies that $|L_S-L_T|$ can be replaced with $|R_S-R_T|$ in Proposition~\ref{prop-LB1}. 
There are pairs of trees without equivalent edges for which the lower bound of Proposition~\ref{prop-LB1} is particularly significant. 
For the trees of Figure 4, for example, we have $L_S=c$ and $L_T=n$, hence
$C(S,T)\geq n-c$ where $c$ is an arbitrary integer constant, $1\leq c\leq n-1$. 
In fact there are many ways of building pairs of trees with $|L_S-L_T|=n-c$ for arbitrary values of $c$. In particular for $c=1$ we have:

\begin{prop}\label{prop-LB1n}
There are pairs of trees $S,T$ without equivalent edges for which $C(S,T)\geq n-1$.
\end{prop}

The lower and upper bounds of Propositions~\ref{prop-UBn} and~\ref{prop-LB1n} are tight. A comparable result is unknown for the standard rotation distance $D(S,T)$ where the upper bound $2n-6$ must be compared with the highest known lower bounds $\frac{5}{3}n-4$ or $2n-\Theta(\sqrt{n})$ proved in~\cite{D10}.
It is also worth noting that for the trees of Figure 4 we have $C(S,T)\leq R_S+R_T-2=n-c$ by Proposition~\ref{prop-UB}, matching the lower bound shown in the figure for any value of $c$. 

\begin{figure}
\begin{center}
\includegraphics[scale=0.8]{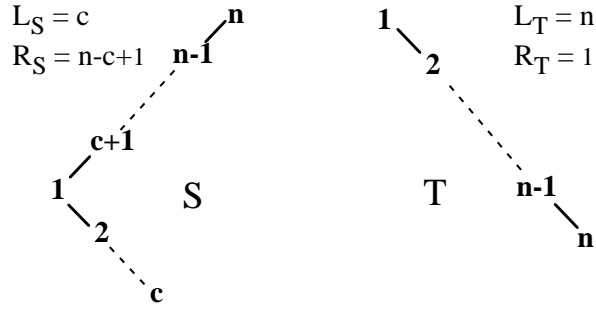}
\caption{Two trees meeting the lower bound $C(S,T)\geq n-c$ by Proposition~\ref{prop-LB1}.}
\end{center}
\end{figure}

\section{Concluding remarks}\label{further}

Standard rotations have been mainly studied in the framework of data organization and computational biology.
When considered as a measure of tree distance, however, different rules may apply to different cases. 
As an example, finding pairs of trees that meet the upper bound of $2n-6$ is outside the realm of balanced search trees where long chains of pointers prevent such a bound to be met~\cite{LP89}. 
Therefore it seems reasonable to investigate other concepts of distance going beyond standard rotations. In this respect we have proposed chain distance here. 
In fact we believe that a possible merit of this note, if any, is stimulating a discussion on the concept of distance between trees.

When looking for alternatives we must observe three main properties, fulfilled both by rotations and chain rotations. First the transformation rely on subtree transfer and on the replacement of a constant number of vertices; then an invariant is maintained in the tree (the infix ordering of vertices in our case); and finally the basic operation requires constant time (rotation or c-rotation require changing three pointers). An alternative to rotations and c-rotations could be moving a single vertex $v$ along a chain with a jump of any length to insert $v$ above one of its ancestors $w$, thus defining a {\em long distance} $L(S,T)$ between two trees. It can be easily seen that also in this case one subtree is relocated,
the infix ordering of the vertices is maintained, and only three pointers are changed. Again $D(S,T)$ is a special case of $L(S,T)$ if only vertex jumps of length one are allowed. 
Upper and lower bounds on the distance should be established in this case. 

In fact, a wealth of possibilities is open.


\end{document}